\documentclass{svjour3}

\begin{document}

\title{Linear and Riccati equations in generating functions for stellar models in
general relativity}
\author{B. V. Ivanov}
\institute{B.V.Ivanov \at Institute for Nuclear Research and Nuclear Energy, \\
Bulgarian Academy of Science, \\
Tzarigradsko Shausse 72, Sofia 1784, Bulgaria \\
\email{boykovi@gmail.com}}
\maketitle

\begin{abstract}
It is shown that the expressions for the tangential pressure, the anisotropy
factor and the radial pressure in the Einstein equations may serve as
generating functions for stellar models. The latter can incorporate an
equation of state when the expression for the energy density is also used.
Other generating functions are based on the condition for the existence of
conformal motion (conformal flatness in particular) and the Karmarkar
condition for embedding class one metrics. In all these cases the equations
are linear first order differential equations for one of the metric
components and Riccati equations for the other. The latter may be always
transformed into second order homogenous linear differential equations.
These conclusions are illustrated by numerous particular examples from the
study of stellar models.
\end{abstract}

\section{Introduction}

Gravitation is governed by the Einstein equations of general relativity in
the simplest case. Unlike the Maxwell equations, which are linear, the
Einstein equations are a system of highly non-linear differential second
order equations in partial derivatives. They are simplified for systems with
symmetry. In astrophysics spherical symmetry is usually used, which reduces
in the static case the differential equations to ordinary ones and the
derivatives are with respect to the radius. The metric is diagonal with just
two components. In canonical comoving coordinates there are three equations
for five unknowns - the two metric potentials and the three diagonal
components of the energy-momentum tensor $T_{ab}$, namely, the energy
density $\mu $, the radial and the tangential pressures $p_r$ and $p_t$.
Thus the fluid is anisotropic, which is backed by arguments for compact
objects with very high density \cite{Ruder} and by a number of other reasons 
\cite{HS}, \cite{BIJTP}.

On one side these equations present expressions for the components of the
energy-momentum tensor. On the other side the metric potentials enter in a
rather involved way as they are obtained from the Ricci tensor and scalar.
The equations remain non-linear for the metric. In 1983, however, Durgapal
and Banerjee \cite{DB} showed that in the perfect fluid case the Einstein
equations are linear of first order for a function of $g_{11}$ and the
equations for $p_t$ and the anisotropy factor are linear of second order for
a function of $g_{00}$. Later, these findings were generalised for
anisotropic fluid \cite{cmone}, but the reason for this simplification
remains unclear.

There is a well-known generating function for stellar solutions with $g_{00}$
and the anisotropy factor as potentials \cite{nd}. It is based on the
equation for the latter, which is a difference of two of the Einstein
equations. Are there other generating functions based on the field
equations? What is the role of an equation of state (EOS)? Does it simplify
or complicate the problem? Are there any common features between the field
equations and other ways to generate a solution, like conformal flatness,
conformal motion or the possibility to embed the spacetime in a flat
5-dimensional spacetime, inspired by string and brane models?

The present paper tries to answer these questions in a systematic way. We
shall not discuss the many conditions for physical viability of the
solution, but concentrate on the mathematical issues that the above
questions invoke and back them with concrete examples from the literature in
the corresponding section.

In Sect. 2 the field equations are given, as well as some characteristics of
the model and the equations for the anisotropy factor, the existence of
conformal motion or flatness in particular, and the Karmarkar condition. In
Sect. 3 we give the three types of differential equations that are used in
the following and list some of their properties. In Sect. 4 a generating
function, based on the expression for the radial pressure is discussed. When
an EOS is imposed, the expression for the energy density is also necessary.
Sect. 5 gives generating function based on the expressions for the
tangential pressure. The well-known generating function, based on the
anisotropy factor, is recovered. In Sect. 6 we discuss in detail generating
functions based on the condition for the existence of conformal motion or
conformal flatness. Sect. 7 deals with spacetimes of embedding class one and
their relation to the previous cases. Sect. 8 gives some arguments for the
appearance of an important structure in the equations. Sect. 9 contains some
discussion.

\section{Field equations and definitions}

The interior of static spherically symmetric stars is described by the
canonical line element 
\begin{equation}
ds^2=e^\nu c^2dt^2-e^\lambda dr^2-r^2\left( d\theta ^2+\sin ^2\theta
d\varphi ^2\right) ,  \label{one}
\end{equation}
where $\lambda $ and $\nu $ depend only on the radial coordinate $r$. The
coordinates are numbered as $x^0=t$, $x^1=r$, $x^2=\theta $ and $x^3=\varphi 
$. The Einstein equations read 
\begin{equation}
k\mu =\frac 1{r^2}-\left( \frac 1{r^2}-\frac{\lambda ^{\prime }}r\right)
e^{-\lambda },  \label{two}
\end{equation}

\begin{equation}
kp_r=-\frac 1{r^2}\left( 1-e^{-\lambda }\right) +\frac{\nu ^{\prime }}%
re^{-\lambda },  \label{three}
\end{equation}
\begin{equation}
kp_t=\frac{e^{-\lambda }}4\left( 2\nu ^{\prime \prime }+\nu ^{\prime 2}+%
\frac{2\nu ^{\prime }}r-\nu ^{\prime }\lambda ^{\prime }-\frac{2\lambda
^{\prime }}r\right) ,  \label{four}
\end{equation}
where $\mu $ is the energy density, $p_r$ is the radial pressure, $p_t$ is
the tangential one, $^{\prime }$ means a radial derivative and 
\begin{equation}
k=\frac{8\pi G}{c^4}.  \label{five}
\end{equation}
Here $G$ is the gravitational constant and $c$ is the speed of light. We use
relativistic units with $G=1=c$. Then $k=8\pi $.

The gravitational mass in a sphere of radius $r$ is given by 
\begin{equation}
m=4\pi \int_0^r\mu \left( \omega \right) \omega ^2d\omega .  \label{six}
\end{equation}
Then Eq. (2) yields 
\begin{equation}
e^{-\lambda }=1-\frac{2m}r.  \label{seven}
\end{equation}

The compactness of the star $u$ is defined by 
\begin{equation}
u=\frac mr  \label{eight}
\end{equation}

The redshift $Z$ depends on $\nu $: 
\begin{equation}
Z\left( r\right) =e^{-\nu /2}-1.  \label{nine}
\end{equation}
The field equations do not contain $\nu $, but its first and second
derivative. It is related to the four-acceleration $a_1$, namely $2a_1=\nu
^{\prime }$.

As a whole, we have three field equations for five unknown functions: $%
\lambda ,\nu ,\mu ,p_r$ and $p_t$. We can choose freely two of them, but the
model will be physically realistic if a number of regularity, matching and
stability requirements are satisfied too.

Different conditions may be imposed on the system of Einstein equations. One
of them is the existence of an equation of state (EOS) $p_r=f\left( \mu
\right) $.

It is useful to introduce the anisotropic factor $\Delta =p_t-p_r$. It
measures the anisotropy of the fluid. Eqs. (3, 4) give 
\begin{equation}
-8\pi \Delta =e^{-\lambda }\left( -\frac{\nu ^{\prime \prime }}2-\frac{\nu
^{\prime 2}}4+\frac{\nu ^{\prime }}{2r}+\frac 1{r^2}\right) +e^{-\lambda }%
\frac{\lambda ^{\prime }}2\left( \frac{\nu ^{\prime }}2+\frac 1r\right)
-\frac 1{r^2}.  \label{ten}
\end{equation}
When $\Delta =0$ the fluid becomes perfect and all pressures are equal.

Another requirement is conformally flat spacetime. It happens when its Weyl
tensor vanishes. One can encompass this in the requirement that a Killing
vector $\mathbf{K}$ exists. Then the following equation has to be satisfied 
\begin{equation}
L_{\mathbf{K}}g_{ab}=2\psi g_{ab},  \label{eleven}
\end{equation}
where $L_{\mathbf{K}}$ is the Lie derivative operator and $\psi \left(
t,r\right) $ is the conformal factor. This implies the equation \cite{cmone} 
\begin{equation}
2\nu ^{\prime \prime }+\nu ^{\prime 2}=\nu ^{\prime }\lambda ^{\prime }+%
\frac{2\nu ^{\prime }}r-\frac{2\lambda ^{\prime }}r+\frac 4{r^2}\left(
1+s\right) e^\lambda -\frac 4{r^2},  \label{twelve}
\end{equation}
where $s$ is a constant of integration. The spacetime is conformally flat
when $s=0$.

In recent years spacetimes, which are embeddings of class one, have been
widely discussed. They can be embedded in a 5-dimensional flat spacetime.
This requires a condition between the components of the Riemann tensor \cite
{Karmarkar} 
\begin{equation}
R_{1010}R_{2323}-R_{1212}R_{3030}=R_{1220}R_{1330}.  \label{thirteen}
\end{equation}
It transforms into a differential equation for $\lambda $ and $\nu $: 
\begin{equation}
2\frac{\nu ^{\prime \prime }}{\nu ^{\prime }}+\nu ^{\prime }=\frac{\lambda
^{\prime }e^\lambda }{e^\lambda -1}.  \label{fourteen}
\end{equation}

\section{Types of equations}

We shall show that Eqs (2, 3, 4, 10, 12) are linear with respect to $%
y=e^{-\lambda }$ while Eq (14) is linear for $y=e^\lambda $. Eq (2) does not
contain $a_1$, while Eq (3) gives an expression for it. The others belong to
the types of equations, discussed below with respect to $a_1$. They may be
transformed into linear equations for $u=e^{\nu /2}$. Therefore we give some
properties of these types of equations \cite{PZ}. Except for the constants $%
C $ and $n$, the other letters designate functions of $r$.

1) Linear equation. It reads 
\begin{equation}
gy^{\prime }=f_1y+f_0.  \label{fifteen}
\end{equation}
It is integrable in quadratures and its general solution is 
\begin{equation}
y=Ce^F+e^F\int e^{-F}\frac{f_0}gdr,\quad F=\int \frac{f_1}gdr.
\label{sixteen}
\end{equation}

2) Bernoulli equation. It reads 
\begin{equation}
gy^{\prime }=f_1y+f_ny^n.  \label{seventeen}
\end{equation}
It turns into a linear equation for $w=y^{1-n}$%
\begin{equation}
gw^{\prime }=\left( 1-n\right) f_1w+\left( 1-n\right) f_n  \label{eighteen}
\end{equation}
and is soluble in general. Using (16) its solution becomes 
\begin{equation}
y^{1-n}=Ce^F+\left( 1-n\right) e^F\int e^{-F}\frac{f_n}gdr,\quad F=\left(
1-n\right) \int \frac{f_1}gdr.  \label{nineteen}
\end{equation}

3) Riccati equation. It is given by 
\begin{equation}
gy^{\prime }=f_2y^2+f_1y+f_0  \label{tw}
\end{equation}
and has no general solution. When $f_2=0$ it turns into a linear equation.
When $f_0=0$ it becomes a Bernoulli equation with $n=2$, so that $1/y$
satisfies a linear equation. Every Riccati equation may be transformed into
a canonical form with $f_1=0$. There is a general solution for the Riccati
equation when one particular solution $y_0$ is known: 
\begin{equation}
y=y_0+F\left( C-\int F\frac{f_2}gdr\right) ^{-1},\quad F=\exp \int \left(
2f_2y_0+f_1\right) \frac{dr}g.  \label{twone}
\end{equation}
This formula simplifies when two or more particular solutions are known.

This equation may be transformed into a second-order homogenous linear
equation for $u$ when the following substitution is made 
\begin{equation}
u=\exp \left( -\int \frac{f_2}gydr\right) .  \label{twtwo}
\end{equation}
Namely, Eq. (20) becomes 
\begin{equation}
g^2f_2u^{\prime \prime }+g\left[ f_2g^{\prime }-gf_2^{\prime }-f_1f_2\right]
u^{\prime }+f_0f_2^2u=0.  \label{twthree}
\end{equation}
In the case $f_2=-g$ Eqs. (22, 23) simplify considerably 
\begin{equation}
u=\exp \int ydr,  \label{twfour}
\end{equation}
\begin{equation}
f_2u^{\prime \prime }+f_1u^{\prime }+f_0u=0.  \label{twfive}
\end{equation}
The substitution $y=u^{\prime }/u$ leads back to Eq (20).

\section{The energy density and the radial pressure}

Eq (2) for the energy density does not contain $a_1$. It is linear with
respect to $y=e^{-\lambda }$ and can be written as 
\begin{equation}
ry^{\prime }=-y+1-8\pi \mu r^2.  \label{twsix}
\end{equation}
When solved, it yields Eq (7) with the mass given by Eq (6). Eq (7) becomes 
\begin{equation}
y=1-\frac{2m}r.  \label{twseven}
\end{equation}
Any equation, linear in $y$ may be transformed into an equation, linear in $%
m $ with the use of the above formula.

Eq (3) for the radial pressure may be written as 
\begin{equation}
8\pi p_rr^2=\left( 2a_1r+1\right) y-1.  \label{tweight}
\end{equation}
It may be regarded as an expression for $p_r$ or $y$%
\begin{equation}
y=\frac{8\pi p_rr^2+1}{2ra_1+1}  \label{twnine}
\end{equation}
or $a_1$%
\begin{equation}
2a_1=\nu ^{\prime }=\frac{8\pi p_rr^2+1-y}{ry}.  \label{thirty}
\end{equation}
The potential $\nu $ is found by a simple quadrature.

Thus, Eq (28), which contains $p_r$, $y$ and $a_1$, is the simplest
generating function for any of them, when the other two are known. In some
papers ansatze for $y$ and $p_r$ were chosen \cite{lpone}, \cite{lptwo}, 
\cite{lpthree}, \cite{lpfour}. One may also choose $p_r$ and $\mu $, because
most of the regularity conditions are imposed on them, and find after that $%
y $ from Eqs (6, 27). Then $\lambda $ and $\nu $ are found, which give the
remaining characteristics of the model \cite{rhoradone}, \cite{rhoradtwo}, 
\cite{rhoradthree}, \cite{rhoradfour}, \cite{flone}, \cite{fltwo}, \cite
{flthree}.

An EOS can be incorporated in this scheme, $p_r=f\left( \mu \right) $ or 
\begin{equation}
2rya_1=1-y+8\pi r^2f\left( -\frac{ry^{\prime }+y-1}{8\pi r^2}\right) ,
\label{thone}
\end{equation}
which follows from Eqs (26, 28). Obviously, the resulting equation is not
linear in $y$ in general, but still may be solvable by choosing an ansatz
for $y$. Anyway, its an expression for $a_1$ in terms of $y$ and is a
relation between the metric potentials. Quadratic EOS $p_r=a\mu ^2+b\mu +c$
have been used in \cite{Qone}, \cite{Qtwo}, \cite{Qthree}, \cite{Qfour}, 
\cite{Qfive}. The polytropic EOS $p_r=\kappa \mu ^{1+1/N}$ with $\kappa $ a
constant and $N$ the polytropic index was studied in \cite{EOSone}, \cite
{EOStwo}. The modified Van der Vaals EOS $p_r=\alpha \mu ^2+\frac{\gamma \mu 
}{1+\beta \mu }$ was assumed in \cite{EOSthree} and \cite{EOSfour} ($%
a,b,c,\kappa ,N,\alpha ,\beta $ are constants). A nonlinear EOS in the
framework of colour superconductivity was used in \cite{EOSfive}, together
with a given mass function or energy density profile. One can find $y$ from
them using Eq (7) and then the other characteristics.

A special case is the linear EOS (LEOS) $p_r=a\mu -b$ with constant $0\leq
a\leq 1$ and the bag constant $b\geq 0$, which includes also the case $p_r=0$%
. Eq (31) becomes 
\begin{equation}
2rya_1=\left( a+1\right) \left( 1-y\right) -ray^{\prime }-8\pi br^2.
\label{thtwo}
\end{equation}
It is also a linear equation for $y$ 
\begin{equation}
ray^{\prime }=-\left( 2ra_1+a+1\right) y+a+1-8\pi br^2.  \label{ththree}
\end{equation}
Eq (16) gives in this case a singular $e^F$ for $r=0$, hence $C=0$. Then the
solution is 
\begin{equation}
y=\frac{\int \left( a+1-8\pi br^2\right) \left( re^\nu \right) ^{1/a}dr}{%
a\left( r^{a+1}e^\nu \right) ^{1/a}}.  \label{thfour}
\end{equation}
In the literature $y$ is given and $a_1$ is found from Eq (32) \cite{LEOSone}%
, \cite{LEOStwo}, \cite{LEOSthree}, \cite{LEOSfour}, \cite{LEOSfive}, \cite
{LEOSsix}, \cite{LEOSseven}, \cite{LEOSeight}, \cite{LEOSnine}, \cite
{LEOSten}. Due to Eqs (6, 7) one finds quickly $y$ form $\mu $ or $m$.
Solutions with LEOS and known $\mu $ are given in \cite{Lrhoone}, \cite
{Lrhotwo}, \cite{Lrhothree} and \cite{Lrhofour}. Solutions with LEOS and
known $m$ are given in \cite{Lmone}, \cite{Lmtwo}, \cite{Lmthree}, \cite
{Lmfour}.

\section{The tangential pressure and the anisotropic factor}

Eq (4) is an expression for $p_t$ and can be written as a linear equation
for $y$ 
\begin{equation}
\frac 12\left( a_1+\frac 1r\right) y^{\prime }=-\left( a_1^{\prime }+a_1^2+%
\frac{a_1}r\right) y+8\pi p_t.  \label{thfive}
\end{equation}
Its solution from Eq (16) reads 
\begin{equation}
y=e^F\left( C+16\pi \int ze^\nu e^{2\int \frac{dr}{r^2z}}p_tdr\right) ,
\label{thsix}
\end{equation}
where 
\begin{equation}
a_1+\frac 1r=\frac{\nu ^{\prime }}2+\frac 1r\equiv z,  \label{thseven}
\end{equation}
\begin{equation}
e^F=z^{-2}e^{-\nu }e^{-2\int \frac{dr}{r^2z}}  \label{theight}
\end{equation}
Due to Eq (27) it is also a linear equation with respect to the mass.

Eq (4) is the only Einstein equation that is also a Riccati equation for $%
a_1 $%
\begin{equation}
ya_1^{\prime }=-ya_1^2-\left( \frac yr+\frac{y^{\prime }}2\right) a_1-\frac{%
y^{\prime }}{2r}+8\pi p_t  \label{thnine}
\end{equation}
and may be solved for particular choices of $y$ and $p_t$. It can be
transformed into a linear second order homogenous differential equation
following Eqs (24, 25) 
\begin{equation}
yu^{\prime \prime }+\left( \frac yr+\frac{y^{\prime }}2\right) u^{\prime
}+\left( \frac{y^{\prime }}{2r}-8\pi p_t\right) u=0,  \label{forty}
\end{equation}
where 
\begin{equation}
u=e^{\nu /2}.  \label{foone}
\end{equation}
Sometimes it may be solved easier than the original Riccati equation, since
many special functions are defined by such equations. It remains in the same
time linear (and integrable) first order equation for $y=e^{-\lambda }$ or $%
m $. It can be called a double linear equation. This equation was found in 
\cite{DB}, where it is divided by $u$ and the variable $r^2$ is used instead
of $r$. This is not necessary.

Thus, like $p_r$, the expression (4) for $p_t$ may be considered as a
generating function for stellar models, when two of the quantities $p_t$, $y$
(or $m$) and $a_1$ are known. It simplifies in the case $p_t=0$ as seen from
Eq (35).

The generating functions based on $\Delta $ are found in a similar way. Eq
(10) is linear with respect to $y$ (or $m$) and may be rewritten as 
\begin{equation}
\left( a_1+\frac 1r\right) y^{\prime }=-2\left( a_1^{\prime }+a_1^2-\frac{a_1%
}r-\frac 1{r^2}\right) y-2\left( \frac 1{r^2}-8\pi \Delta \right) .
\label{fotwo}
\end{equation}
After some transformations it becomes 
\begin{equation}
y^{\prime }=-2\left( \frac{z^{\prime }}z+z-\frac 3r+\frac 2{r^2z}\right)
y-\frac 2z\left( \frac 1{r^2}-8\pi \Delta \right) .  \label{fothree}
\end{equation}
This is exactly Eq (8) from \cite{nd}, which may be integrated. The result 
\begin{equation}
y=r^6z^{-2}e^{-\int \left( \frac 4{r^2z}+2z\right) dr}\left( C-2\int
r^{-8}z\left( 1-8\pi \Delta r^2\right) e^{\int \left( \frac
4{r^2z}+2z\right) dr}dr\right)  \label{fofour}
\end{equation}
is the same as in \cite{nd}, when the different definition of their $\Delta $
is taken into account. The generating potentials are $\Delta $ and $z$, the
second, due to Eq (37), is equivalent to $a_1$. This generating function
encompasses the important case of perfect fluid $\Delta =0$, when the
formula (44) has been found earlier \cite{ndzero}. These potentials have
been used in \cite{ndzeroa}, \cite{ndone}, \cite{ndtwo}, \cite{ndthree}.
Models with perfect fluid have been discussed recently \cite{ndfour}, \cite
{ndfive}, \cite{ndfivea}.

However, Eq (42) is also a Riccati one for $a_1$, the Riccati structure $%
a_1^{\prime }+a_1^2$ being brought in $\Delta $ by $p_t$. It can be written
as 
\begin{equation}
2ya_1^{\prime }=-2ya_1^2+\left( \frac{2y}r-y^{\prime }\right) a_1+\frac{%
2y-2-ry^{\prime }}{r^2}+16\pi \Delta  \label{fofive}
\end{equation}
and solved for particular $\Delta $ and $y$. Finally, it can be linearised
following Eqs (24, 25) into 
\begin{equation}
-2yu^{\prime \prime }+\left( \frac{2y}r-y^{\prime }\right) u^{\prime
}+\left( \frac{2y-2-ry^{\prime }}{r^2}+16\pi \Delta \right) u=0,
\label{fosix}
\end{equation}
where $u$ is given by Eq (41). Thus, once again, Eq (46) is doubly linear,
like Eq (40). In total, Eq (10) is a generating function for stellar models,
when two of the quantities $\Delta $, $y$ (or $m$) and $a_1$ are known. The
differential equations for $y$ and $u$ are linear. Eq (46) has been used in 
\cite{ldone}, \cite{ldtwo}, \cite{ldthree}, \cite{ldfour}, \cite{ldfoura}, 
\cite{ldfourb}, \cite{ldfive}, \cite{ldsix}, \cite{ldseven}, \cite{ldeight}, 
\cite{ldnine}.

It is well-known that the Newtonian polytropes satisfy the non-linear
Lane-Emden equation for perfect or anisotropic fluids \cite{dpone}. It is
integrable for just a few values of its parameter. This equation has an
extension in general relativity for isotropic and anisotropic fluids, based
on the TOV equation, which is even more intricate \cite{dptwo}, \cite
{dpthree}, \cite{ppolyone}, \cite{ppolytwo}. We have seen in Sect. 4 that
fluids with EOS are treated in a much simpler way when Eqs (2, 3) are taken
as generating functions. Thus Newtonian mechanics, where the metric is flat,
seems to be more complicated than general relativity for building of stellar
models.

The simplest way to generate solutions is to choose independently the two
generating potentials $\lambda $ and $\nu $. This works when the fluid is
anisotropic in general. Such models were discussed in \cite{metone}, \cite
{mettwo}, \cite{metthree}, \cite{metfour}, \cite{metfive}, \cite{metsix}, 
\cite{metseven}, \cite{meteight}, \cite{metnine}.

Some important stellar models require a relation between the two metric
potentials, reducing the generating functions to one. For example this is
the case of perfect fluid, when in Eq. (10) $\Delta =0$. One may say that
then the metric ''admits'' a perfect fluid as a source. Models with EOS also
lead to such relations, see Eqs (31, 32, 33). Two similar examples are
discussed in the next two sections.

\section{Spacetimes admitting conformal motion}

The metric potentials of such spacetimes satisfy Eq (12). In \cite{cmone}
this equation is solved by a series of transformations and adding terms.
Somewhat surprisingly, it is also a linear equation in $y$ (or $m$) and a
Riccati equation for $a_1$. It can be written as 
\begin{equation}
\left( \frac 1r-a_1\right) y^{\prime }=2\left( a_1^{\prime }+a_1^2-\frac{a_1}%
r+\frac 1{r^2}\right) y-\frac{2\left( 1+s\right) }{r^2}  \label{foseven}
\end{equation}
or 
\begin{equation}
2ya_1^{\prime }=-2ya_1^2+\left( \frac{2y}r-y^{\prime }\right) a_1+\frac{%
y^{\prime }}r+\frac{2\left( 1+s\right) -2y}{r^2}.  \label{foeight}
\end{equation}

Let us discuss the linear equation first, which is always integrable. In
this particular case the integrals are expressed by simple functions.

The field equations give an expression for $\nu ^{\prime }$%
\begin{equation}
\nu ^{\prime }=\frac{krp_r+2m/r^2}{1-2m/r}.  \label{fonine}
\end{equation}
The regularity properties show that $a_1\rightarrow 0$ when $r\rightarrow 0$
and therefore $1/r-a_1$ is positive there. Then Eq (16) gives\quad 
\begin{equation}
e^F=\frac 1{e^\nu \left( \frac 1r-a_1\right) ^2}  \label{fifty}
\end{equation}
and 
\begin{equation}
y=e^F\left[ C+\left( 1+s\right) \frac{e^\nu }{r^2}\right] ,  \label{fione}
\end{equation}
where $C$ is an arbitrary constant of integration. Passing to $\lambda $ and 
$\nu $ we get 
\begin{equation}
\frac{e^\lambda }{r^2}=\frac{\left( \frac 1r-\nu ^{\prime }\right) ^2}{%
1+s+Cr^2e^{-\nu }}.  \label{fitwo}
\end{equation}
This formula expresses $\lambda $ in terms of $\nu $. Near the centre of the
star the l.h.s. goes to $+\infty $, while the r.h.s. goes to $\infty /\left(
1+s\right) $. Hence, $1+s\geq 0$. This includes the conformally flat case $%
s=0$.

Eq (52) may be inverted as an expression of $\nu $ through $\lambda $. For
this purpose we integrate it and obtain 
\begin{equation}
2\int \frac{e^{\lambda /2}}rdr=\int \frac{dX}{X\sqrt{1+s+CX}},\quad
X=r^2e^{-\nu }.  \label{fithree}
\end{equation}
The right integral is given in elementary functions and when $1+s>0$, 
\begin{equation}
\exp \left( C_1+2\sqrt{1+s}\int \frac{e^{\lambda /2}}rdr\right) =Z,\quad
Z=\mid \frac{\left( 1+s+CX\right) ^{1/2}-\left( 1+s\right) ^{1/2}}{\left(
1+s+CX\right) ^{1/2}+\left( 1+s\right) ^{1/2}}\mid .  \label{fifour}
\end{equation}
Obviously $Z<1$.

When $C<0$ one has 
\begin{equation}
\frac{CX}{1+s}=\frac{-4Z}{\left( 1+Z\right) ^2}  \label{fifive}
\end{equation}
and 
\begin{equation}
e^{\nu /2}=\sqrt{\frac{-C}{1+s}}r\cosh \left( \sqrt{1+s}\int \frac{%
e^{\lambda /2}}rdr+\frac{C_1}2\right) ,  \label{fisix}
\end{equation}
which is the solution used in \cite{wthree}. Comparing it to the solution in 
\cite{cmone}, we have 
\begin{equation}
Ae^x+Be^{-x}=c_1\cosh \left( x+c_2\right) ,  \label{fiseven}
\end{equation}
where $x$ denotes the bracket in Eq (56). This is true only when $A,B$ have
the same sign.

When $C>0$ one has 
\begin{equation}
\frac{CX}{1+s}=\frac{4Z}{\left( 1-Z\right) ^2}  \label{fieight}
\end{equation}
and 
\begin{equation}
e^{\nu /2}=-\sqrt{\frac C{1+s}}r\sinh \left( \sqrt{1+s}\int \frac{e^{\lambda
/2}}rdr+\frac{C_1}2\right) .  \label{finine}
\end{equation}
This solution corresponds to the case when $A$ and $B$ differ in sign.

When $1+s=0$ Eq (52) gives 
\begin{equation}
e^{\nu /2}=-\sqrt{C}r\left( \int \frac{e^{\lambda /2}}rdr+\frac{C_1}2\right)
,  \label{sixty}
\end{equation}
which coincides with the solution in \cite{cmone}.

Thus, Eqs (52, 56, 59, 60) may be considered as generating functions for
spacetimes admitting conformal motion. Such solutions have been studied also
in \cite{cmtwo}, \cite{cmthree}, \cite{cmfour}, \cite{cmfive}, \cite{cmsix}, 
\cite{cmseven}, \cite{cmeight}. Conformally flat solutions are discussed in 
\cite{wzero}, \cite{wone}, \cite{wtwo}, \cite{wthree}, \cite{wfour}, \cite
{wfive}.

Eq (47) may be written as a Riccati equation for $a_1$%
\begin{equation}
2ya_1^{\prime }=-2ya_1^2+\left( \frac{2y}r-y^{\prime }\right) a_1+\frac{%
y^{\prime }}r+\frac{2\left( 1+s\right) -2y}{r^2}.  \label{sione}
\end{equation}
Once again $g=-f_2$ in Eq (20), so it may be transformed into a linear
equation, analogous to Eq (25) 
\begin{equation}
-2yu^{\prime \prime }+\left( \frac{2y}r-y^{\prime }\right) u^{\prime
}+\left( \frac{y^{\prime }}r+\frac{2\left( 1+k\right) -2y}{r^2}\right) u=0,
\label{sitwo}
\end{equation}
where $u$ is given by Eq (24).

\section{Spacetimes of embedding class one}

The Karmarkar condition Eq (14) may be written as 
\begin{equation}
a_1^{\prime }=-a_1^2+\left[ \ln \left( q-1\right) \right] ^{\prime }\frac{a_1%
}2,  \label{sithree}
\end{equation}
where $q=1/y=e^\lambda .$ The Riccati structure from the previous sections
emerges once again, but there is no free term. Thus, the would be Riccati
equation becomes a Bernoulli one (see Eq (17)) with $n=2$ and is integrable.
According to Eq (18) it is a linear equation with respect to $w=1/a_1$%
\begin{equation}
w^{\prime }=-\left[ \ln \left( q-1\right) \right] ^{\prime }\frac w2+1.
\label{sifour}
\end{equation}

On the other hand, Eq (63) may be rewritten as a linear equation for $q$%
\begin{equation}
\frac{a_1}2q^{\prime }=\left( a_1^{\prime }+a_1^2\right) q-\left(
a_1^{\prime }+a_1^2\right) .  \label{sifive}
\end{equation}
Replacing in it $y=1/q$ we obtain a Bernoulli equation for $y$ with $n=2$%
\begin{equation}
-\frac{a_1}2y^{\prime }=\left( a_1^{\prime }+a_1^2\right) y-\left(
a_1^{\prime }+a_1^2\right) y^2.  \label{sisix}
\end{equation}
We may call Eq (14) a double Bernoulli for $a_1$ and $y$ and a double linear
for $w$ and $q$. All these equations are solvable. Their integration may be
done directly, without using the general formulas and we obtain the
well-known results 
\begin{equation}
e^\lambda =C\nu ^{\prime 2}e^\nu +1,  \label{siseven}
\end{equation}
\begin{equation}
e^\nu =\left( A+B\int \sqrt{e^\lambda -1}dr\right) ^2.  \label{sieight}
\end{equation}
where $A,B,C$ are integration constants.

We can use Eq (7) to express $\lambda $ through the mass $m$ and then Eq
(68) to express $\nu $ through $m$. Hence, the line element (1) depends only
on $m$, which through Eq (6) is given by the energy density $\mu $, i.e. a
component of the energy-momentum tensor. This was done recently in \cite
{kzero}, using the formalism of complex scalars in the static case, albeit
in different notation. The same operation may be repeated in the case with
conformal motion, using Eqs (7, 56, 59, 60). However, the problem is not
simplified.

Embedding class one solutions have been studied extensively in the last
years. Usually an ansatz for $e^\lambda $ is given, but $e^\nu $ is used
instead sometimes. There are solutions with polynomial metric \cite{kone}, 
\cite{konea}, \cite{koneb}, \cite{ktwo}, \cite{kthree}, \cite{kfour}, \cite
{kfive}, \cite{ksix}, \cite{kseven}, \cite{keight}, rational metric \cite
{keighta}, \cite{keightb}, \cite{knine}, \cite{kten}, \cite{keleven}, \cite
{ktwelve}, \cite{kthirteen}, \cite{kfourteen}, \cite{kfifteen}, exponential
one \cite{ksixteen}, \cite{kseventeen}, with trigonometric functions \cite
{keighteen}, \cite{knineteen}, \cite{ktwenty}, \cite{ktwone}, \cite{ktwtwo}, 
\cite{ktwthree}, \cite{ktwthreea}, \cite{ktwfour}, \cite{ktwfive}, \cite
{ktwsix}, \cite{ktwseven} hyperbolic \cite{ktwsevena}, \cite{ktwsevenb}, 
\cite{ktweight}, \cite{ktwnine}, \cite{kthirty}. Special functions are also
used like the hypergeometric one \cite{kthirtya} and the error function \cite
{kthone}.

\section{On the origin of the Riccati structure}

The presence of the combination $a_1^{\prime }+a_1^2$ in all the generating
functions, discussed in the previous sections, requires some explanation.
Let us start with the Riemann tensor for spherically symmetric static
spacetimes. Its only components, written in the notation of the present work
are \cite{ktwthree} 
\begin{equation}
R_{1010}=-e^\nu \left( a_1^{\prime }+a_1^2-\frac{\lambda ^{\prime }}%
2a_1\right) ,  \label{sinine}
\end{equation}
\begin{equation}
R_{3030}=-e^{\nu -\lambda }a_1r\sin ^2\theta ,  \label{seventy}
\end{equation}
\begin{equation}
R_{1212}=-\frac r2\lambda ^{\prime },  \label{seone}
\end{equation}
\begin{equation}
R_{2323}=-\left( 1-e^{-\lambda }\right) r^2\sin ^2\theta .  \label{setwo}
\end{equation}
Only $R_{1010}$ contains the Riccati structure and it enters the Karmarkar
condition (13). This explains its appearance there. The r.h.s. of Eq (13)
vanishes, while all terms on the l.h.s. contain $a_1,a_1^{\prime }$ or $%
a_1^2 $. Therefore there is no free term and the Riccati equation is
truncated to a Bernoulli one.

The field equations follow from the well-known formula 
\begin{equation}
G_{ab}\equiv R_{ab}-\frac 12Rg_{ab}=kT_{ab},  \label{sethree}
\end{equation}
where $R_{ab}$ is the Ricci tensor, $R$ is the Ricci scalar, $G_{ab}$ is the
Einstein tensor, $g_{ab}$ is the metric and $T_{ab}$ is the energy-momentum
tensor. We have the contractions 
\begin{equation}
R_{ab}=R_{\ acb}^c,\quad R=R_{\ c}^c.  \label{sefour}
\end{equation}
Both $R_{00}$ and $R_{11}$ include the term $R_{1010}$ and hence, the
Riccati structure. However, they enter also $R$ and this structure cancels
in $G_{00}$ and $G_{11}$. Thus $\mu $ and $p_r$ do not contain it, but $p_t$
does. It also enters $\Delta $ through the tangential pressure.

Finally, let us discuss spacetimes with conformal motion \cite{cmone}. The
Weyl tensor $C_{abcd}$ has in general a very complicated expression. In the static spherically
symmetric case all its components are proportional to an object $\Gamma $,
which includes $a_1^{\prime }+a_1^2$. Eq\ (11) can be transformed into 
\begin{equation}
L_{\mathbf{K}}C_{\ bcd}^a=0,  \label{sefive}
\end{equation}
which is written as 
\begin{equation}
\Gamma =\frac{e^{2\lambda }}{r^2}s.  \label{sesix}
\end{equation}
This equation is equivalent to Eq (12) or (47).

Thus, we have traced formally the appearance of the Riccati structure,
starting with $R_{1010}$.

\section{Discussion}

In this paper we have studied the existence of generating functions, giving
solutions for stellar models. The general results are backed by concrete
examples from the literature. We have shown that the field equations for the
radial pressure, the tangential pressure and the anisotropy factor may be
used as generating functions when two of the three characteristics of the
model in any of them are known. The equation for $\Delta $ was used in \cite
{nd} to obtain $\lambda $ when $\nu $ and $\Delta $ are given. It becomes a
generating function for perfect fluid models when $\Delta =0$. We have shown
that Eq (4) for the tangential pressure can play a similar role. The
simplest generating function is based on Eq (3) for the radial pressure,
which is an expression for $p_r$, $a_1=\nu ^{\prime }/2$, or $y=e^{-\lambda
} $ without solving any equations. Eq (2) for the energy density cannot be
used as a generating function, because it does not contain $\nu $. However,
when there is an EOS, the combination of Eqs (2) and (3) works as a
generating function, producing a relation between the two metric potentials.

This approach is greatly simplified, because it turns out that the field
equations are linear first order differential equations for $y$ and Riccati
equations for the four-acceleration $a_1$. The first are always integrable
in quadratures, while the second are integrable in many particular cases.
There is a formula for their general solution when a particular solution is
known. They also may be transformed into linear homogenous differential
equations of second order for $u=e^{\nu /2},$ thus being the ''missing
link'' between the original form of the Einstein equations and their linear
version, which appears out of nowhere in \cite{DB}. Eq (7) shows that the
mass $m$ also satisfies a linear equation and may replace $y$.

Of course, the simplest generating potentials are $\lambda $ and $\nu $.
There are physical reasons that sometimes impose a relation between them.
This happens when an EOS exists.

A second important case is that of spacetimes admitting conformal motion
(which includes the case of conformal flatness of the metric). The
surprising fact is that this relation is also a linear differential equation
for $y$ or $u$ and a Riccati one for $a_1$.

A third well-known example are spacetimes of embedding class one, obeying
the Karmarkar condition, which are studied intensively in the past few
years. Here there is a minor difference - the relation is a linear equation
for $1/y$ and a Bernoulli equation for $y$, which is also completely
integrable. Furthermore, it is a Bernoulli equation with quadratic term for $%
a_1$. The Riccati structure, discussed above, is still present but there is
no free term.

In the previous section we have traced the source of the Riccati structure $%
a_1^{\prime }+a_1^2$ and found that it comes from one of the components of
the Riemann tensor, which survives contractions to the Ricci tensor and the
Ricci scalar, enters the Weyl tensor and its Lie derivative or comes
directly from the Karmarkar condition. This also answers the question about
the appearance of the linear homogenous equation for $u$. We could not
clarify why all these equations, on which generating functions are based,
are linear of first order for $y$ (or $1/y$ in the embedding problem).

\end{document}